\documentclass[final,1p,times]{elsarticle} 
\usepackage{graphicx} 
\usepackage{amssymb} 
\usepackage{amsthm} 
\usepackage{lineno} 

\journal{Nuclear Physics A} 

\newcommand{\hm}{\hat{m}}
\newcommand{\VEV}[1]{\left\langle #1\right\rangle}

\begin{document} 

\begin{frontmatter} 


\title{The HotQCD Equation of State}

\author{R.A. Soltz$^{a}$ for the HotQCD collaboration}

\address[a]{Lawrence Livermore National Laboratory, 
7000 East Ave.,
Livermore, CA 94550,  USA}

\begin{abstract} 
We present results from recent calculations of the QCD equation of state by the HotQCD Collaboration and review the implications for hydrodynamic modeling.  The equation of state of QCD at zero baryon density was calculated on a lattice of dimensions $32^3 \times 8$ with $m_l = 0.1~m_s$ (corresponding to a pion mass of $\sim$220~ MeV) using two improved staggered fermion actions, p4 and asqtad.  Calculations were performed along lines of constant physics using more than 100M cpu-hours on BG/L supercomputers at LLNL, NYBlue, and SDSC.  We present parameterizations of the equation of state suitable for input into hydrodynamics models of heavy ion collisions.
\end{abstract} 

\end{frontmatter} 


\section{Calculating the Equation of State with Lattice QCD}
Lattice Quantum Chromodynamics (LQCD) is the only viable method for calculating the equation of state for Quantum Chromodynamics in the vicinity of the transition between hadrons and the quark gluon plasma.    As such, LQCD provides an essential input for modeling and thereby understanding the physics of relativistic heavy ion collisions.  Using BG/L class supercomputers at LLNL, BNL, and SDSC, the HotQCD collaboration has performed caculations of the equation of state and developed parameterizations suitable for input into hydrodynamic models.  The essential results for these calcualtions are summarized in this proceedings.  Additional details and insights are found in Bazavov et al.~\cite{Bazavov:2009zn}.

As described in the seminal work of Wilson~\cite{Wilson:1974sk}
and Creutz~\cite{Creutz:1980zw}, the partition function for the QCD action can be calculated numerically on a discrete Euclidean lattice.  In this form, the gluon action becomes a sum over all closed four-link loops (plaquettes) that can be evaluated with finite difference approximations.  The full theory of QCD is recovered in the continuum limit, in which the lattice spacing is reduced to zero.  Both the p4 and asqtad gauge actions are Symanzik improved, which means that they incorporate a six-link term that improves the convergence to $O(a^2)$.  Both p4 and asqtad {\em stagger} the fermion spinors on odd-even lattice sites, which preserves a discrete chiral symmetry (i.e. exact chiral symmery is recovered in the continuum limit).  Both actions also include $O(a^2)$ improvements, but they differ in their approach to further reducing cutoff effects.  The p4 actions uses fat link smearing~\cite{Cheng:2007jq}, whereas the asqtad action uses tadpole coefficients~\cite{Bernard:2006nj}.

\begin{figure}[ht]
\centering
\includegraphics[width=0.46\textwidth]{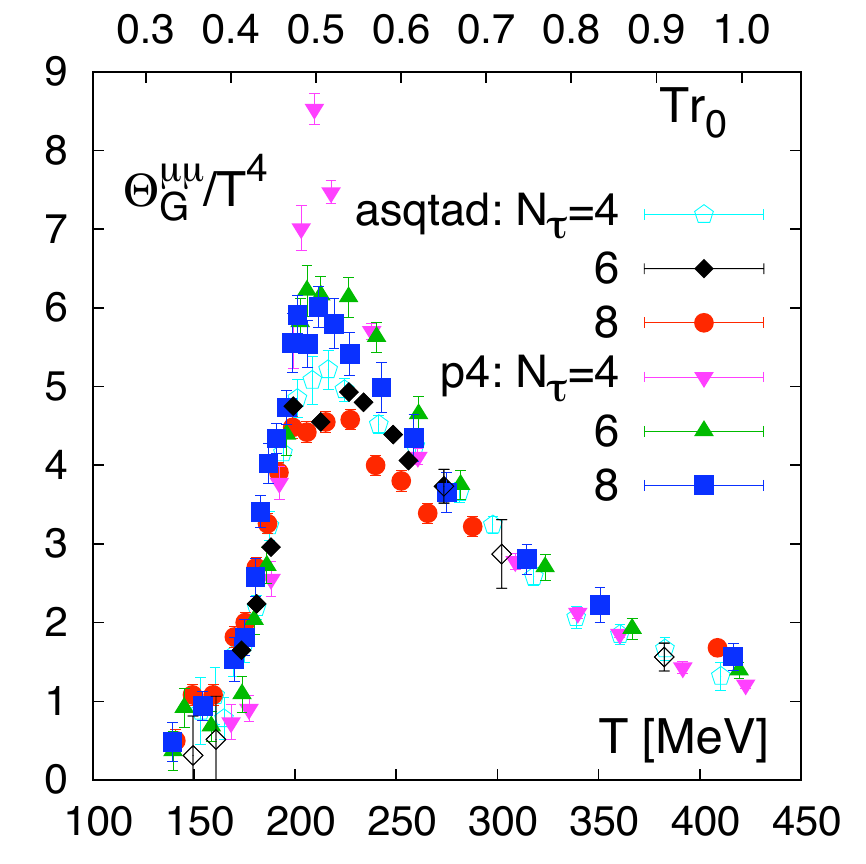}
\includegraphics[width=0.47\textwidth]{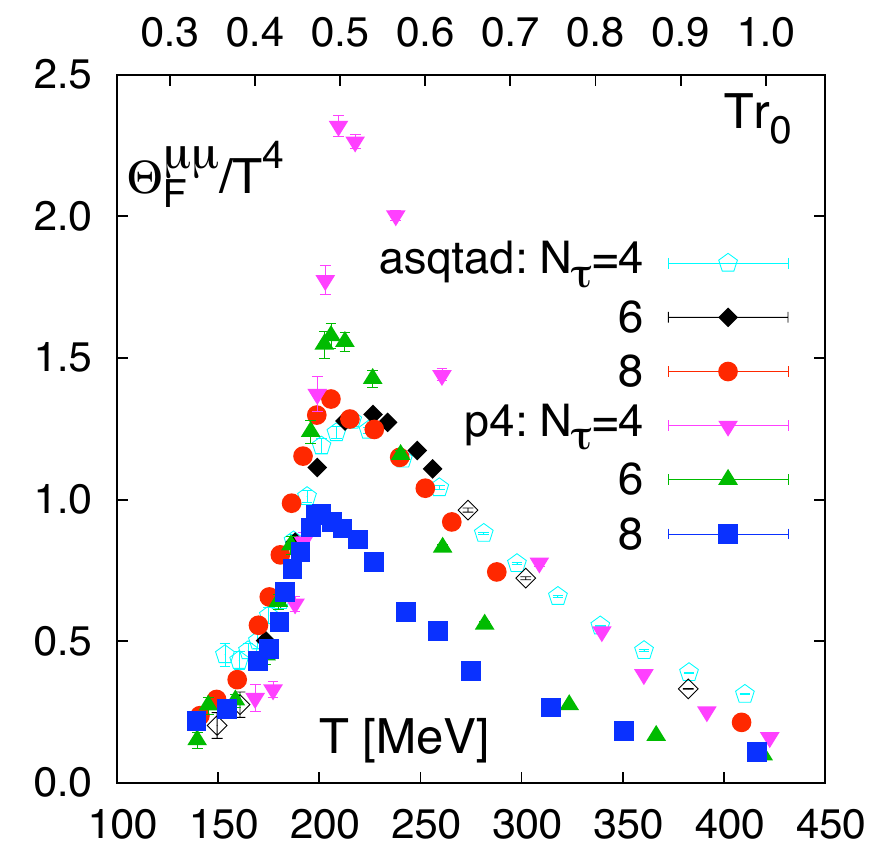}
\caption[]{Contributions the trace anomaly from gauge (left) and fermion (right).}
\label{theta_GF}
\end{figure}
The runs were performed along lines of constant physics, meaning that the bare quark masses are adjusted to maintain a constant value of the pion mass for a light quark mass set to one tenth the value of the strange quark mass.  The temperature is determined by 
determining the distance $r_0$ at which the slope of the heavy  quark potential equals $1.65/r_0^2$.   We use $r_0=0.469(7)$~fm in this analysis~\cite{Cheng:2007jq}.  The thermodynamic quantities were calculated by averaging over approximately 15k trajectories for 24 temperature values for p4, and 17 temperatures for asqtad, with a corresponding zero temperature runs of approximately 5k trajectories.  A complete listing of run statistics, parameters, and thermodynamic quantities is given in the appendices of~\cite{Bazavov:2009zn}.

The pressure and energy density are derived from the calculation of the interaction measure or trace anomaly, $\Theta^{\mu\mu}(T)/T^4=(\epsilon-3p)/T^4$, which measures the deviation from the conformal limit.  The pressure is obtained by integrating above a temperature, $T_0$, for which the trace anomlay is close to zero.
\begin{equation}
\frac{p(T)}{T^4} - \frac{p(T_0)}{T_0^4} = \int_{T_0}^{T} {\rm d}T'
\frac{1}{T'^5} \Theta^{\mu\mu} (T') \;\; .
\label{pres}
\end{equation}

The expression for the trace anomaly from the fermion and gluon components is,
\begin{eqnarray}
\frac{\Theta^{\mu\mu}_F(T)}{T^4}  
&=&-R_\beta R_m N_\tau^4  
\left( 2 \hm_l \Delta\VEV{\bar \psi \psi}_l
        + \hm_s \Delta\VEV{\bar \psi \psi}_s \right) \; , \\
\frac{\Theta^{\mu\mu}_G(T)}{T^4}  
&=&  R_\beta N_\tau^4 \left( \Delta\VEV{s_G} -R_u 
\left( 
6 \beta'_{\rm rt} \Delta\VEV{R} + 4 \beta'_{\rm pg} \Delta\VEV{C}
+\frac{1}{4\beta}  
\Delta\VEV{{\rm Tr}\left(\left(2 D_l^{-1}+D_s^{-1}\right)
\frac{d M}{d u_0} \right) }\right)
\right)
 \; . 
\label{e3pfg} 
\end{eqnarray}
Here we use the notation $\Delta\VEV{X} = \VEV{X}_0 -\VEV{X}_\tau$ to denote subtraction of the corresponding zero temperature values.
$R_u$ is related to the derivative of the tadpole coefficient with respect to the gauge coupling and is used in the asqtad calculations; this parameter is set to zero for p4.

\section{Equation of State Results}
\begin{figure}[th]
\centering
\includegraphics[width=0.6\textwidth]{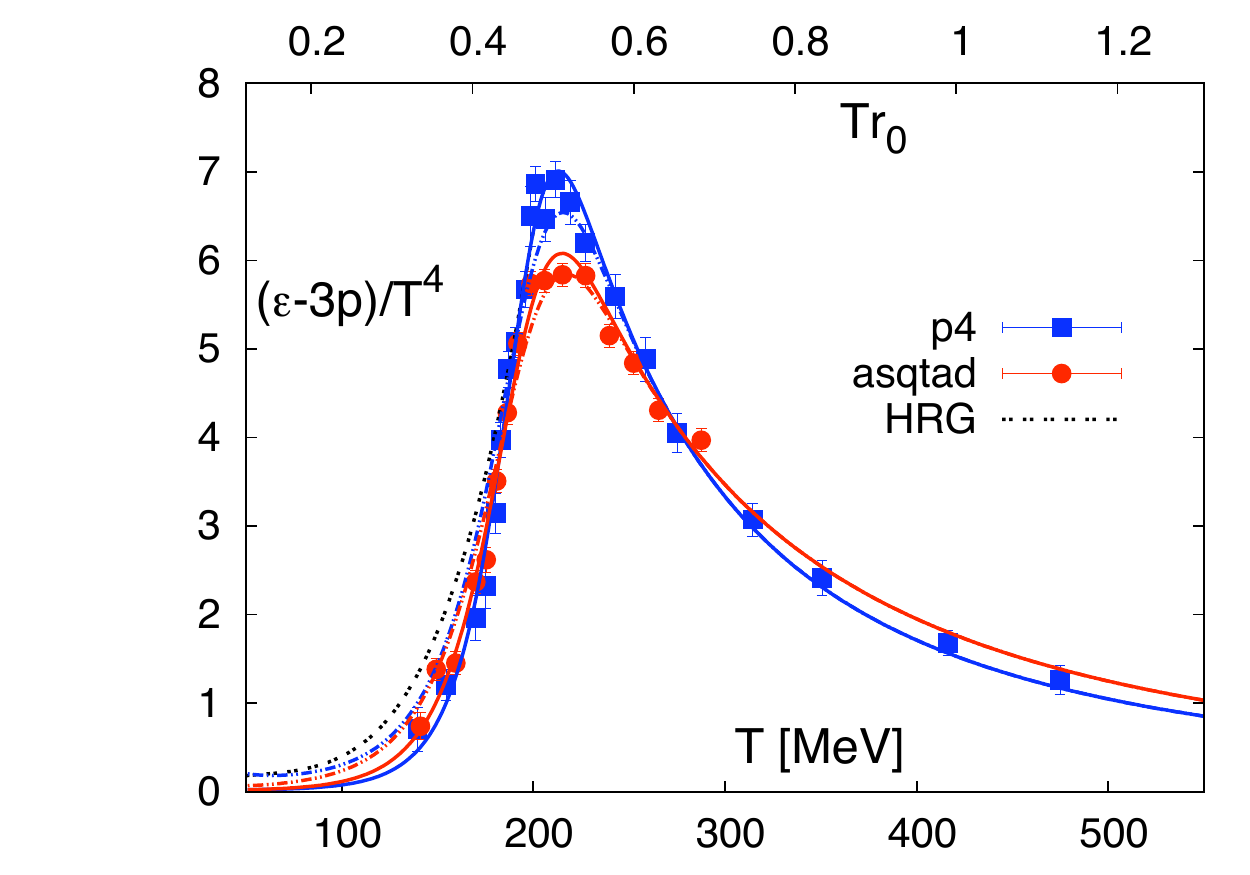}
\caption[]{Paramerized fits to the trace anomaly for p4 and asqtad calculations on lattices with $N_\tau=8$.}
\label{em3p_fits}
\end{figure}

%
%
The contributions from the gauge and fermion fields to the trace anomaly are shown separately in Figure~\ref{theta_GF}.  The dominant contribution is from the gluon action (left), although the cutoff effects appear to be larger for the fermion action (right).  Note that this figure also includes results for $N_\tau$=4,6 lattices.  The full trace anomaly calculations for p4 and asqtad are shown in Figure~\ref{em3p_fits}.  The two actions show good agreement in the low and high temperature regions, although the peak is lower for asqtad than for p4.  Figure~\ref{em3p_fits} also shows the parameterized fits of the lattice results, and fits to lattice results joined to hadron resonance gas calculations below 130~MeV.

\section{Parameterizations for Hydrodynamics}

\begin{figure}[th]
\centering
\includegraphics[height=1.9in]{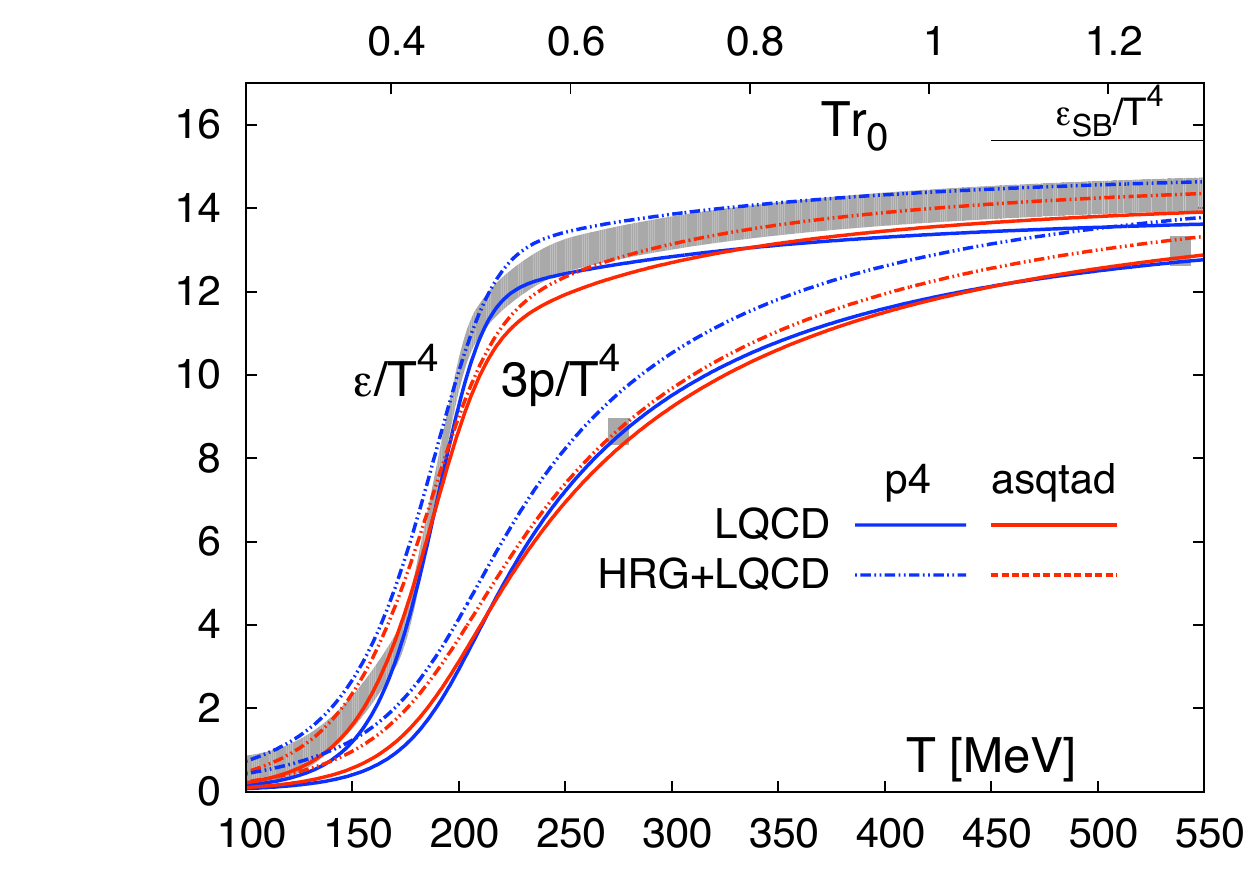}
\includegraphics[height=1.9in]{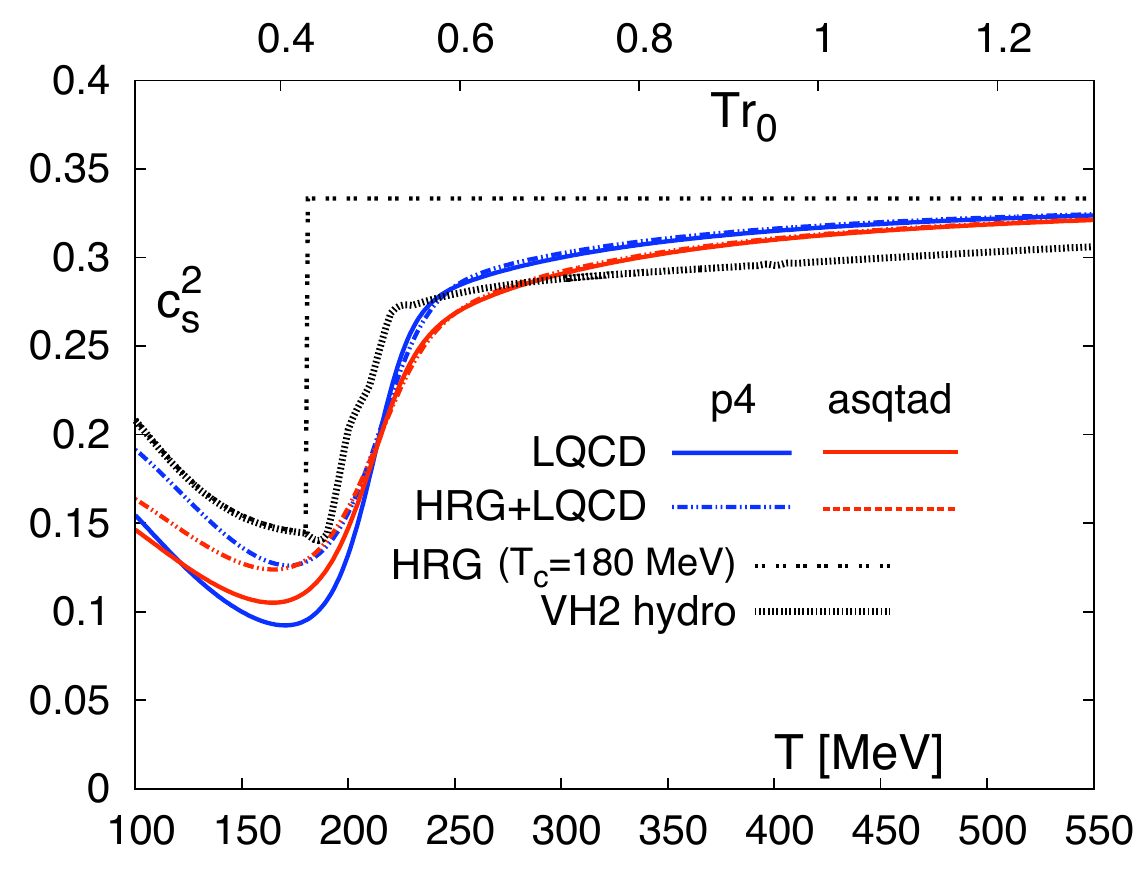}
\caption[]{Energy density and pressure (left) and square of the speed of sound for parameterizations of the trace anomaly given by Eq.~\ref{eq:simplefit}.}
\label{eos_compare}
\end{figure}

To facilitate the use of the HotQCD equation of state in hydrodynamic calculations we provide a simple parameterization for the trace anomaly in Eq.~(\ref{eq:simplefit}).
\begin{equation}
\frac{\epsilon -3p}{T^4} = 
\left( 1 - \frac{1}{[1+e^{(T-c_1)/c_2)}]^{2}} \right)
\left(  \frac{d_2}{T^2}+  \frac{d_4}{T^4} \right)
\label{eq:simplefit} 
\end{equation}
The parameters for these fits are listed in Table~\ref{tab:simplefit} for both the lattice fits and fits to the lattice and HRG results below 130~MeV.  Additional details of the fitting procedure and systematic variations are given in Appendix C of ~\cite{Bazavov:2009zn}.
\begin{table}[h]
\begin{center}
\vspace{0.3cm}
\begin{tabular}{|l|c|c|c|c|c|}
\hline
Data & $d_2$ [GeV]$^2$ & $d_4$ [GeV]$^4$ & $c_1$ [GeV] & $c_2$ [GeV] &
$\chi^2$/dof \\
\hline
            p4  & 0.24(2) &  0.0054(17) &  0.2038(6)   &  0.0136(4)   
&  26.7/19 \\
HRG+p4  & 0.24(2) &  0.0054(17) &  0.2073(6)   &  0.0172(3)   &  -- \\
\hline
              asq   & 0.312(5)  & 0.00  &  0.2024(6)  &  0.0162(4) & 
34.4/14 \\
HRG+asqtad & 0.312(5) & 0.00 &  0.2048(6)  &  0.0188(4) & -- \\
\hline
\end{tabular}
\caption{Parameter values for fits of Eq.~(\ref{eq:simplefit}) to trace
anomaly data for p4 and asqtad, and for data combined with HRG calculations.}
\label{tab:simplefit}
\end{center}
\end{table}
The pressure, energy density, and speed of sound squared are shown in Figure~\ref{eos_compare}.  The pressure was derived from Eq.~\ref{eq:simplefit} by numerical integration, beginning at a temperature of 50~MeV, the approximate minimum for this functional form.  The shaded region for the energy density indicates the error associated with the linear vs. spline interpolation method described in~\cite{Bazavov:2009zn}, and the shaded region near the pressure shows the contribution from a hadron resonance gas calculation for the region below 100~MeV.  Figure~\ref{eos_compare} also shows the speed of sound squared for a hadronic gas with a first order phase transition (double-dotted) as well as the default lattice inspired equation of state  used by the 2D+1 viscous hydrodynamic model of Luzum and Romatschke~\cite{Luzum:2008cw}.

We have presented new results on the equation of state of QCD with
a strange quark mass chosen close to its physical value and two degenerate light
quarks with one tenth of the strange quark mass. A comparison of 
calculations performed with the p4 and asqtad staggered fermion 
discretization schemes shows that both actions lead to a consistent
picture for the temperature dependence of bulk thermodynamic observables.
We have provided a simple parameterization for the equation of state 
suitable for insertion into hydrodynamic models of heavy ion collisions.
Additional discussion of deconfinement and chiral symmetry
restoration aspects 
of QCD thermodynamics for the calculations on lattices with
$N_\tau=8$ described herein may be found in~\cite{Bazavov:2009zn}. 


\section*{Acknowledgments} 
This work was supported in part by contracts DE-AC02-98CH10886, DE-AC52-07NA27344,
DE-FG02-92ER40699, DE-FG02-91ER-40628, DE-FG02-91ER-40661, DE-KA-14-01-02, 
DE-FG02-04ER-41298 with the U.S. Department of Energy, and NSF grants PHY08-57333, 
PHY07-57035, PHY07-57333 and PHY07-03296,
the Bundesministerium f\"ur Bildung und Forschung under grant
06BI401, the Gesellschaft
f\"ur Schwerionenforschung under grant BILAER and the Deutsche
Forschungsgemeinschaft under grant GRK 881.

\end{document}